\PassOptionsToPackage{amsthm}{newtxmath}
\documentclass{optica-article}

\journal{opticajournal}
\articletype{Research Article}

\usepackage{cleveref}
\Crefname{equation}{Eq.}{Eqs.}
\Crefname{figure}{Fig.}{Figs.}
\Crefname{tabular}{Tab.}{Tabs.}
\Crefname{section}{Sec.}{Secs.}
\usepackage{amsmath}
\usepackage{amsfonts}
\usepackage{mathtools}
\usepackage{bm}
\usepackage{multirow}
\usepackage{booktabs}
\usepackage{wrapfig}
\usepackage[labelfont=bf,font=small]{caption}
\usepackage{subcaption}

\newcommand{\dv}[1]{\,\mathrm{d} #1}
\DeclareMathOperator{\hess}{H}
\DeclareMathOperator{\Tr}{Tr}
\DeclarePairedDelimiter{\parens}{\lparen}{\rparen}
\DeclarePairedDelimiter{\abs}{\lvert}{\rvert}
\DeclarePairedDelimiter{\norm}{\lVert}{\rVert}
\makeatletter
\usepackage{doi}
\usepackage{algorithmicx}

\newtheoremstyle{yly}
                {3pt}
                {3pt}
                {\rmfamily}
                {0pt}
                {\bfseries}
                {.}
                {1em}
                {}
\theoremstyle{yly}
\newtheorem{problem}{Problem}

\renewcommand{\vec}[1]{\ensuremath{\bm{\mathrm{#1}}}}

\begin{document}

\title{High-fidelity holographic beam shaping with optimal transport and phase diversity}

\author{Hunter Swan,\authormark{1,*} Andrii Torchylo,\authormark{1} \mbox{Michael J.\ {Van de Graaff}},\authormark{1} Jan Rudolph,\authormark{1,2} and \mbox{Jason M.\ Hogan}\authormark{1,+}}

\address{\authormark{1}Department of Physics, Stanford University, Stanford, California 94305, USA}
\address{\authormark{2}Fermi National Accelerator Laboratory, Batavia, Illinois 60510, USA}

\email{\authormark{*}orswan@stanford.edu}
\email{\authormark{+}hogan@stanford.edu}

\begin{abstract*} 
A phase-only spatial light modulator (SLM) provides a powerful way to shape laser beams into arbitrary intensity patterns, but at the cost of a hard computational problem of determining an appropriate SLM phase.  Here we show that optimal transport methods can generate approximate solutions to this problem that serve as excellent initializations for iterative phase retrieval algorithms, yielding vortex-free solutions with superior accuracy and efficiency.  Additionally, we show that analogous algorithms can be used to measure the intensity and phase of the input beam incident upon the SLM via phase diversity imaging.  These techniques furnish flexible and convenient solutions to the computational challenges of beam shaping with an SLM.
\end{abstract*}

\section{Introduction}

The phase-only spatial light modulator (SLM) has in recent years become an ubiquitous tool for laser beam shaping~\cite{rosalesshape}, with applications to diverse fields such as laser ablation and materials processing~\cite{lutz2021efficient,kuang2015ultrafast}, electron beam shaping~\cite{maxson2015adaptive}, laser projection displays~\cite{chang2014full}, and optical trapping of ultra-cold atomic gasses~\cite{Pasienski:08,harte2014conjugate,Schroff2023}.  SLMs have the potential to generate essentially arbitrary laser intensity patterns with high diffraction efficiency and fast update speeds.  However, realizing this potential requires the solution of two generally challenging problems: 
\begin{itemize}
\item \textit{Phase generation}: Finding the appropriate SLM phase to shape a given incident light field (``input beam'') into a desired output intensity.
\item \textit{Beam estimation}: Determining the amplitude and phase of the input beam.
\end{itemize}
\noindent
Most previous efforts have been devoted to developing iterative algorithms for solving the phase generation problem. These computational holography methods inevitably entail some tradeoff between accuracy and diffraction efficiency of the output beam intensity. In this work we develop new methods for solving both of the above problems using ideas from optimal transport and phase diversity imaging, simultaneously improving accuracy and efficiency.

Optimal transport (OT) is a mathematical framework for finding the optimal manner of moving one probability distribution into another, subject to some cost for the moving process~\cite{Villani2021,Villani2009}.  The origins of OT are closely tied to economic problems of resource allocation, but the subject has found numerous applications to problems including image segmentation~\cite{papadakis2015optimal}, stochastic control of dynamical systems~\cite{chen2021optimal}, and electron density functional theory~\cite{Buttazzo}.  In this work we use OT to find an optimal mapping of light from the input intensity distribution to the target output beam intensity distribution, similar to OT methods for caustic design~\cite{meyron2018light, glimm2003optical, wang2004design, schwartzburg2014high, gutierrez2009refractor}.  

Our approach to phase generation connects electromagnetic wave propagation to one of the fundamental equations of OT theory, the Monge-Ampere equation, and exploits this connection to build an OT-based algorithm for finding approximate solutions (“OT solutions”) to the phase generation problem.  This algorithm may be viewed as a generalization of classical geometrical beam shaping formulas for symmetric beam shapes~\cite{Dickey} to arbitrary input and output intensity profiles, and it has several technical advantages over existing methods: The resulting solutions represent an \textit{unwrapped} phase, are guaranteed to be free of phase vortices~\cite{SENTHILKUMARAN200543}, can be interpolated to different coordinate meshes, and have high diffraction efficiency.  Moreover, our implementation requires essentially no hand tuning (e.g. hyperparameter tuning) and only modest computational resources.

Being approximate, OT solutions typically should be refined by some other phase generation algorithm, such as Gerchberg-Saxton (GS)~\cite{gerchberg1972practical}, Mixed-Region Amplitude Freedom (MRAF)~\cite{Pasienski:08}, or Cost Function Minimization (CFM)~\cite{harte2014conjugate}.  In this work, we have implemented such a refinement procedure with GS and MRAF.  These solutions remain vortex-free in regions of appreciable input intensity and feature an accuracy and efficiency surpassing that produced by the same refining algorithms initialized by other means.  In most state-of-the-art phase generation methods, much care is required in choosing a good initialization~\cite{Pasienski:08,Schroff2023,pasienski2011transport}.  Even for methods which can avoid optical vortex formation from arbitrary initialization (such as forced annihilation~\cite{SENTHILKUMARAN200543} or careful cost function tuning~\cite{harte2014conjugate}), there are typically penalties for doing so, such as increased solution roughness and slower convergence.  We thus suggest that OT solutions can serve as universal initial guesses for phase generation algorithms.  

The accuracy of phase generation is limited by the accuracy of input beam estimation.  To solve the latter problem, we employ a version of phase diversity imaging, which is a technique for measuring phase and intensity of a light field using multiple images of the beam under known perturbations~\cite{Gonsalves1982}.  These perturbations can be achieved in various ways, such as defocusing the imaging camera or applying a phase mask before the imaging plane.  This technique was famously used to characterize aberrations in the Hubble telescope main mirror, and it has found subsequent application to adaptive optics control and exoplanet imaging~\cite{gonsalves2018phase}. 

We develop a model of phase diversity imaging which has much in common with our model of phase generation, including a connection to OT theory.  We introduce and test algorithms for approximating the input beam similar to our phase generation algorithm.  We also describe and test an iterative Fourier transform (IFT) algorithm strictly analogous to Gerchberg-Saxton and show it produces highly accurate input beam estimates.  These techniques have advantages over existing beam estimation methods~\cite{vcivzmar2010situ, Zupancic:16,demars2020single}: They require no additional hardware in the SLM setup; they can recover both phase and intensity of the input beam; they require only a few calibration images; and the spatial resolution of the resulting beam estimate can approach the pixel size of the SLM.  

Details and derivations for each section are presented in parallel sections of the Supplement.  Figure data in this work were generated using our SLMTools Julia package~\cite{SLMTools}.

\section{Setup and mathematical formulation}
\label{sec:math}

\begin{figure}[t]
\centering
\includegraphics[width=0.65\columnwidth]{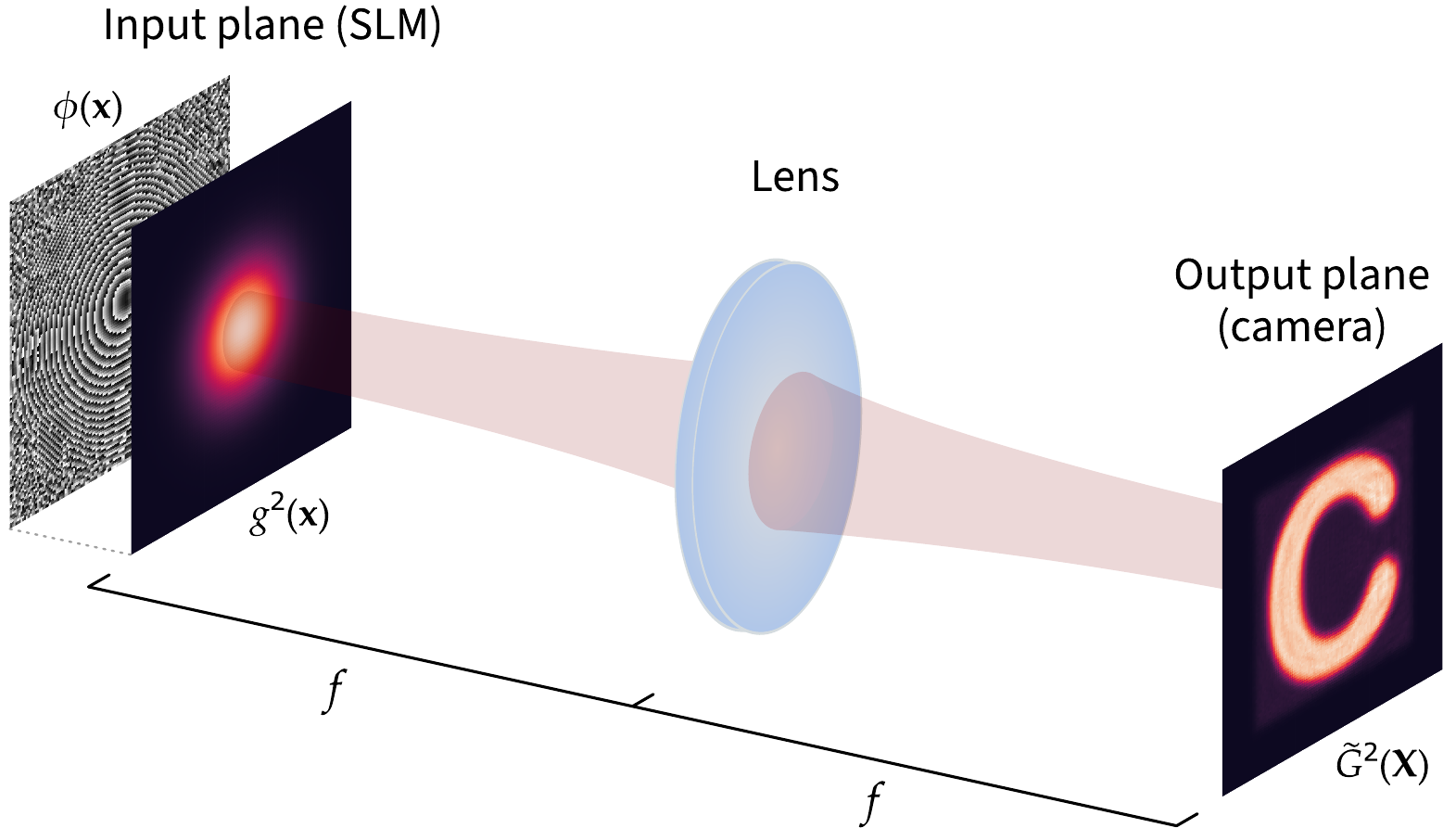}
\caption{Model optical system.  An input laser beam with intensity $g^2(\vec{x})$ is reflected off an SLM with applied phase $\phi(\vec{x})$, passes through a lens of focal length $f$ at distance $f$ from the SLM, and is then imaged on the output (camera) plane at distance $f$ from the lens, with output intensity $\tilde{G}^2(\vec{X})$.}
\label{fig:setup}
\end{figure}

The optical system we consider is shown in \Cref{fig:setup}.  We assume the paraxial limit, that the lens is thin and aberration-free, that the tilt of the SLM is negligible, and that the pixelation and discretization of phase/intensity levels of the SLM and camera are negligible. 

We use lower case letters to refer to quantities in the plane of the SLM and upper case letters to denote corresponding quantities in the plane of the camera.  We non-dimensionalize all distances with length scale $\sqrt{f\lambda}$, where $f$ is the lens focal length and $\lambda$ the light wavelength.  Under the aforementioned approximations, the electric field amplitude at the plane of the SLM $\mathbf{a}(\mathbf{x})$ is related to that at the plane of the camera $\mathbf A (\mathbf X)$ by 
\begin{align}
\mathbf A (\mathbf{X}) & = \iint_{\mathbb{R}^2} \mathbf a (\mathbf{x}) \,e^{-2\pi i \, \mathbf{x} \cdot \mathbf{X}} \dv{\mathbf{x}} \nonumber \\ 
& = \mathcal{F}\!\left[\mathbf a (\mathbf{x})\right]\!(\mathbf{X}),
\label{eq:propagator}
\end{align}
where $\mathcal{F}$ denotes the Fourier transform (unitary convention).

We assume a laser beam linearly polarized in direction $\mathbf{\hat n}$ and denote the beam moduli in the SLM and camera planes by $g(\mathbf{x}) \coloneq \left|\mathbf a (\mathbf x ) \cdot \mathbf{\hat n}\right|$ and $G(\mathbf{X}) \coloneq \left|\mathbf A (\mathbf X) \cdot \mathbf{\hat n}\right|$.  The total power in the electric field $\mathbf a (\mathbf x)$ is given by $\norm{g}_2^2 = \iint_{\mathbb{R}^2} g^2(\mathbf x) \dv{\mathbf x}$, where $\norm{\cdot}_2$ denotes the $L^2$ norm.  We denote the phase of the input beam by $2\pi\,\psi(\mathbf x)$, such that $\mathbf a (\mathbf x) = g(\mathbf x)\, e^{2\pi i \, \psi(\mathbf x)} \mathbf{\hat n}$.  We will often refer to a quantity like $\psi(\mathbf x)$ as a phase, though it is measured in cycles rather than radians.

\subsection{The phase generation problem}
Suppose that we are given a known input beam modulus $g(\mathbf x)$ and target output beam modulus $G(\mathbf X)$, and we wish to find a phase $\phi(\mathbf x)$ so that the realized output beam modulus $\tilde G(\mathbf X) \coloneq \left|\mathcal{F}\left[g(\mathbf{x})\, e^{2\pi i \,\phi(\mathbf x)}\right]\!(\mathbf{X} )\right|$ is equal to $G(\mathbf X)$.  In general, exact solutions to this problem do not exist. For example, a standard result in Fourier analysis says that a function and its Fourier transform cannot both have compact support.  Thus if $g$ and $G$ have compact support, no exact solution $\phi$ exists. We thus relax the requirement for exact equality and formulate phase generation as

\begin{problem}
\label{prob:phase-generation}
Given input beam modulus and target output beam modulus $g,G:\mathbb{R}^2\rightarrow\mathbb{R}_{\geq 0}$ with $\norm{g}_2 = \norm{G}_2$, find a phase function $\phi:\mathbb{R}^2\rightarrow\mathbb{R}$ minimizing 
\begin{equation}
d\left(G(\mathbf{X} ) , \abs*{\mathcal{F}\!\left[g(\mathbf{x})\, e^{2\pi i \, \phi(\mathbf x)}\right]\!(\mathbf{X} )}\right),
\label{eq:prob-1-metric}
\end{equation}
where $d$ is some chosen distance function. 
\end{problem}

The choice of distance $d$ is somewhat arbitrary.  Conventional choices include the $L^2$ distance $d(A,B) \coloneq \norm{A - B}_2$ and an RMS intensity distance defined in \Cref{sec:ot-performance} below.

If a given unwrapped phase $\phi(\mathbf x)$ is convex and well behaved (see Supplement), the Fourier transform in \Cref{eq:propagator} may be estimated using the stationary phase approximation (SPA) as
\begin{equation}
\label{eq:ft-spa-estimate}
\left|\mathcal{F}\!\left[g(\mathbf{x})\, e^{2\pi i \, \phi(\mathbf x)}\right]\!(\mathbf{X} )\right| \approx \frac{g(\mathbf{x})}{\sqrt{\det \hess\phi(\mathbf x)}},
\end{equation}
where $\mathbf x$ satisfies $\nabla\phi(\mathbf x) = \mathbf X$ and $\hess$ is the Hessian operator.  Thus a phase function $\phi$ satisfying
\begin{equation}
G(\mathbf{X} ) = G\parens[\big]{\nabla \phi(\mathbf x)} \approx \frac{g(\mathbf{x})}{\sqrt{\det \hess\phi(\mathbf x)}}
\end{equation}
will provide an approximate solution to Problem~\ref{prob:phase-generation}.  Squaring this relation yields a non-linear partial differential equation,
\begin{equation}
G^2\parens[\big]{\nabla\phi(\mathbf x)} \det \hess\phi(\mathbf x) = g^2(\mathbf x),
\label{eq:MongeAmpere}
\end{equation}
known as the Monge-Ampere equation (MAE).  In \Cref{sec:otpg}, we exploit a connection between optimal transport and the MAE to efficiently solve the latter and thereby get a good estimate of the solution to Problem~\ref{prob:phase-generation}. The phase so generated is convex\cite{de2014monge}, justifying that assumption in Eq. (\ref{eq:ft-spa-estimate}).

In the ray optics perspective, \Cref{eq:MongeAmpere} can be interpreted as simply a condition for local energy conservation.  Rays propagate parallel to the local phase gradient, and the effect of the $2f$ imaging system between the SLM and camera is to send the pencil of rays at point $\mathbf x$ in the SLM plane to a pencil of rays at point $\mathbf X = \nabla \phi (\mathbf x)$ in the camera plane.  The Jacobian $\det \hess \phi$ measures the change in an area element under this mapping.  The MAE thus states that the total power in a pencil of rays is constant as it propagates from the SLM to camera plane. 

\begin{figure*}[t]
    \centering
    \includegraphics[width=0.9\textwidth]{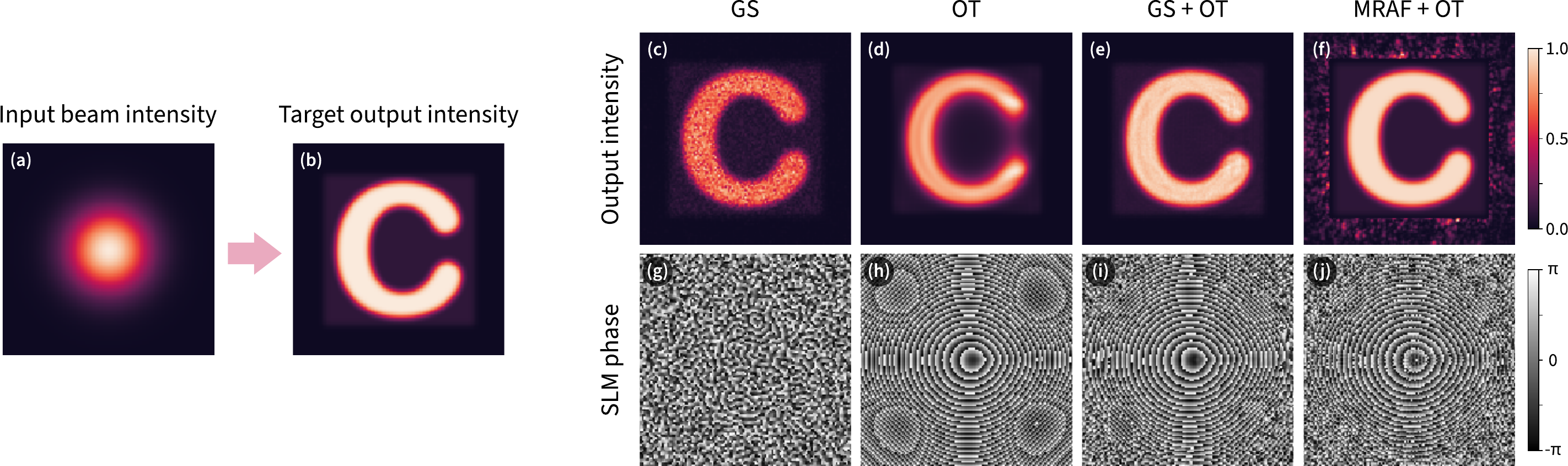}
    \caption{Comparison of phases and output beams from various phase generation algorithms.  All images are $128 \times 128$ pixels.  (a) is the input beam intensity.  (b) is the target output beam intensity. (c-f) are output intensities realized by the phase displayed immediately below. (g) is from GS initialized with a random phase, with RMS error $\epsilon = 13.9\%$ and efficiency $\eta = 99.13\%$. (h) is from OT; $\epsilon = 14.3\%$, $\eta = 99.96\%$. (i) is from GS initialized by OT; $\epsilon =2.58\%$, $\eta = 99.91\%$. (j) is MRAF initialized by OT; $\epsilon = 5.95\times10^{-16}$, $\eta = 85.15\%$.  All iterative algorithms were run for 10,000 iterations.  The MRAF hyperparameter was set by hand to 0.48. A centered $96\times 96$ pixel box was used as the MRAF signal region and the region for computing all efficiencies $\eta$.}
    \label{fig:pg-comparison}
\end{figure*}

\subsection{Phase diversity imaging}
\label{sec:phase-diversity-math}
Phase diversity imaging~\cite{Gonsalves1982} is a reversal of the phase generation problem: Instead of knowing the input beam and finding the phase, we apply several known phases and use the resulting images to determine an unknown input beam.  Specifically, we use $m\geq 2$ quadratic phases $e^{2\pi i \, \alpha_j x^2/2}$, where $\alpha_j\in\mathbb{R}$, $j=1,\dots,m$, and $x^2$ is a shorthand for $\mathbf x \cdot \mathbf x$. We refer to each such phase as a ``diversity phase''.  The use of quadratic phases instead of more general functional forms is related to favorable analytical properties described below.  In an experimental implementation, each diversity phase would also include a linear phase ramp $e^{2\pi i \, \bm{\beta} \cdot \mathbf x}$ to separate the output beam from parasitic undiffracted light (see Supplement).

For each diversity phase $e^{2\pi i \, \alpha_j x^2/2}$ we measure a corresponding output beam modulus
\begin{equation}
G_j(\mathbf X) = \left|\,\mathcal{F}\left[g(\mathbf x)\, e^{2\pi i \parens[\big]{\psi(\mathbf x) + \alpha_j x^2/2}}\right]\!(\mathbf X)\right|,
\end{equation}
where $g(\mathbf x)$ and $\psi(\mathbf x)$ are the unknown input beam modulus and phase.  We refer to $G_j^2$ as a ``diversity image'' and $G_j$ as a ``diversity image modulus''. 
Then the mathematical formulation of phase diversity imaging becomes

\begin{problem}
\label{prob:beam-estimation}
Given coefficients $\alpha_j\in \mathbb{R}$ and diversity image moduli $G_j:\mathbb{R}^2\rightarrow\mathbb{R}_{\geq 0}$ with $\norm[\big]{G_j}_2 = 1$, $j=1,\dots,n$, find $g:\mathbb{R}^2\rightarrow\mathbb{R}_{\geq 0}$, $\psi:\mathbb{R}^2\rightarrow\mathbb{R}$ minimizing 
\begin{equation}
\sum_j  d\left(G_j , \left|\mathcal{F}\!\left[g(\mathbf{x})\, e^{2\pi i \left(\psi(\mathbf x) + {\alpha_j x^2 }/{ 2}\right)}\right]\!(\mathbf X)\right|\right),
\label{eq:pd-metric}
\end{equation}
where $d$ is a chosen distance function.
\end{problem}

For subsequent analysis, we define $\Phi_j(\mathbf X)$ to be the phase associated to diversity modulus $G_j$, so that the electric field of diversity image $j$ is $G_j(\mathbf X)\, e^{2\pi i \,\Phi_j(\mathbf X)}\, \mathbf{\hat n}$.

As with phase generation, under appropriate technical hypotheses (see Supplement) we can apply the SPA to the Fourier transform in \Cref{eq:pd-metric}, yielding the estimates
\begin{align}
G_j(\mathbf X) & \approx \frac{g(\mathbf{x})}{\sqrt{\alpha_j^2 + \alpha_j \Tr \hess \psi(\mathbf{x}) + \det \hess \psi(\mathbf{x})}}, \label{eq:spotlight-pdi-amp} \\
\Phi_j(\mathbf X) & \approx \alpha_j x^2/2 - \mathbf x \cdot \mathbf X,
\label{eq:spotlight-pdi-phase}
\end{align}
where $\mathbf x$ satisfies $\nabla\psi(\mathbf x) + \alpha_j\, \mathbf x = \mathbf X$.  Note that the preceding equation is only valid when $\alpha_j$ is sufficiently large such that the combined phase on the input beam $\psi(\mathbf x) + \alpha_j x^2/2$ is convex or concave.  If we have prior knowledge that the intrinsic phase of the input beam is negligible (e.g.\ if it is well collimated) or if $|\alpha_j|$ is sufficiently large, then Eq.~(\ref{eq:spotlight-pdi-amp}) gives us an immediate estimate for the input beam modulus, 
\begin{equation}
g(\mathbf x) \approx |\alpha_j| \; G_j(\alpha_j \mathbf x),
\label{eq:pd-stationary-phase-interp}
\end{equation}
which says that the input beam modulus is a rescaling of the diversity modulus $G_j$.  Alternatively, from \Cref{eq:spotlight-pdi-phase}, with $\mathbf x \approx \mathbf X/\alpha_j$, 
\begin{equation}
g(\mathbf x) \approx \left|\mathcal{F}^{-1}\!\left[G_j(\mathbf X) \,e^{-2\pi i \, X^2 / (2\alpha_j)}\right]\!(\mathbf x)\right|.
\label{eq:pd-one-shot}
\end{equation}
In practice, \Cref{eq:pd-one-shot} is slightly more convenient than \Cref{eq:pd-stationary-phase-interp}, as discussed in \Cref{sec:one-shot}. 

\begin{figure*}[t]
\includegraphics[width=0.8\textwidth]{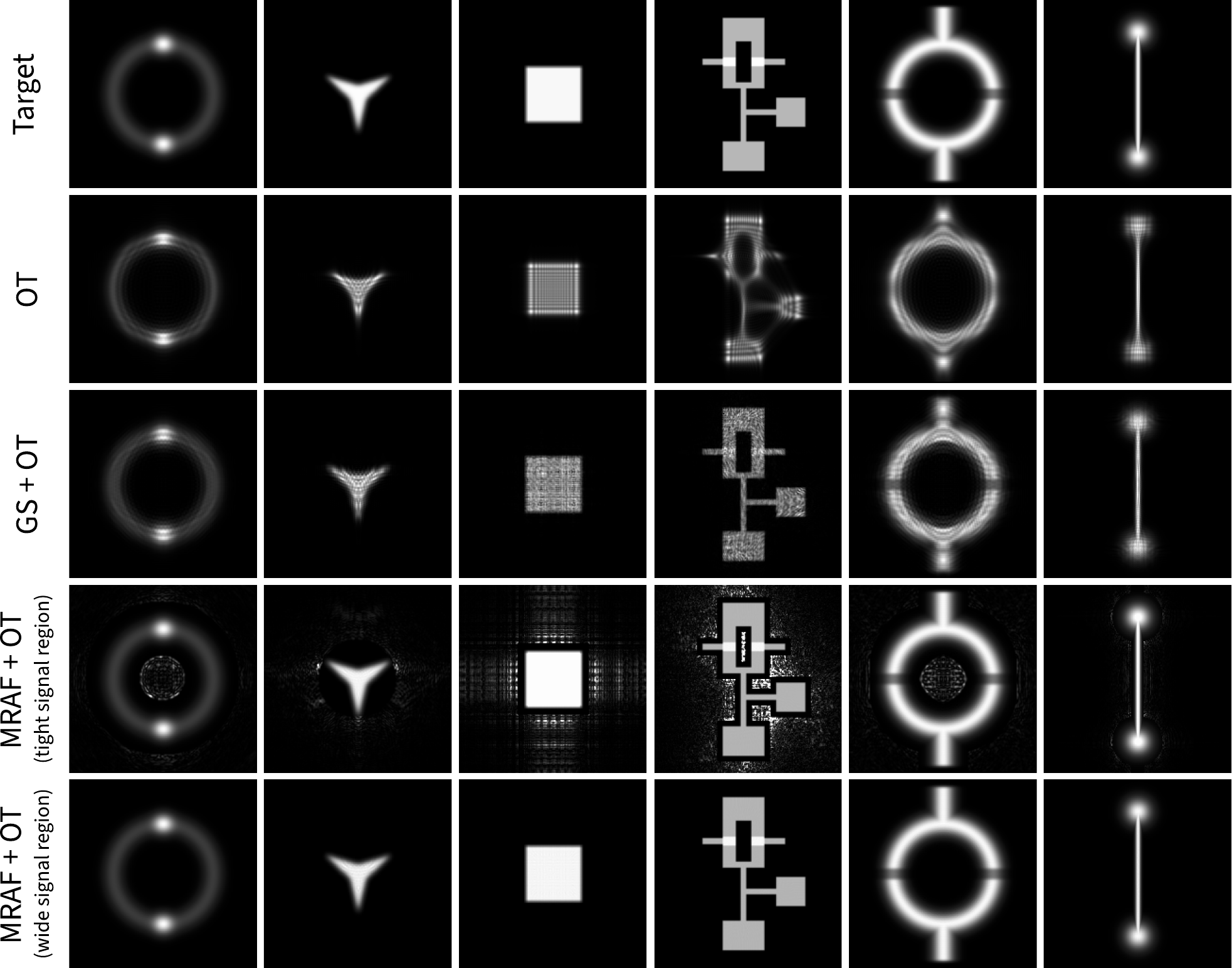}
\centering
\caption{Output beam intensities resulting from various combinations of OT and IFT algorithms.  The top row is a collection of target output beams from Ref.~\cite{Pasienski:08}.  The second row is the output of our OT method with no further refinement.  The third row is the output of the GS algorithm initialized with OT.  The fourth row is the output of the MRAF algorithm with a tight signal region (following Ref.~\cite{Pasienski:08}) and initialized with OT.  The fifth row is MRAF with signal region the entire field of view, again initialized with OT.}
\label{fig:grid-targets}
\end{figure*}

There is a relationship between any two diversity images' electric fields which allows phase generation algorithms to be used for beam estimation.  Namely, by inverse Fourier transforming the field of diversity image $j$, multiplying by the phase $e^{2\pi i (\alpha_k-\alpha_j) x^2 /2}$, and Fourier transforming, one finds
\begin{equation}
\label{eq:pd2shot}
G_k(\mathbf X)\, e^{2\pi i\, \Phi_k(\mathbf X)} = \frac{e^{\frac{2\pi i \,X^2}{2\Delta\alpha}}}{i\Delta\alpha}  \mathcal{F}\left[G_j(\mathbf Y)\, e^{2\pi i \left(\Phi_j(\mathbf Y) + \frac{Y^2}{2\Delta\alpha}\right)}\right]\!\left(\frac{\mathbf X}{\Delta\alpha}\right),
\end{equation}
where $\Delta\alpha \coloneq \alpha_j - \alpha_k$ and the Fourier transform is taken over the variable $\mathbf Y$.  This relationship only holds when quadratic diversity phases are used, and is the reason for that choice.  Apart from rescaling the argument of the Fourier transform and multiplication by known phases, this has the same form as the relationship between the input and output beams in the phase generation problem.  Thus applying SPA yields a Monge-Ampere equation for the phase $\Phi_j(\mathbf X)$,  
\begin{equation}
\label{eq:pdma}
 G_j(\mathbf X)^2 = G_k\parens[\Big]{\Delta\alpha \,\nabla\parens*{\Phi_j(\mathbf X)+\tfrac{X^2}{2\Delta\alpha}}}^2 \det\hess\parens*{\Phi_j(\mathbf X)+\tfrac{X^2}{2\Delta\alpha}}.
\end{equation}
Upon solving this equation for $\Phi_j$, we can determine the unknown input beam electric field $g(\mathbf{x})\, e^{2\pi i\, \psi(\mathbf x)}$ by inverse Fourier transformation of $G_j(\mathbf{X})\, e^{2\pi i \,\Phi_j(\mathbf{X})}$.

\subsection{Discretization}
\label{sec:discretization}

For simulation of the optical system described above, we discretize the SLM and camera planes on rectangular grids with one computational grid point per pixel.  In particular, we ignore subpixel effects such as pixel crosstalk~\cite{moser2019model}. For computational implementation of \Cref{eq:propagator}, we always use sampling grids which are dual in the Fourier sense, and we use a discrete Fourier transform to approximate the continuous Fourier transform~\cite{epstein2005well}. For the optimal transport algorithms of \Cref{sec:otpg,sec:phase-diversity}, this duality constraint is not necessary and any computational grids suffice. See Supplement for further details. 

\begin{table*}[t]
\captionsetup{width=.9\linewidth}
\caption{Performance comparison of GS and MRAF algorithm with and without OT initialization.  $\epsilon$ is the RMS error in percent, with $0$ being optimal.  $\eta$ is the efficiency in percent, with $100$ being optimal.  The target geometries and input beam parameters are identical to those in Ref.~\cite{Pasienski:08} (see Supplement).}\label{table:pg-comparison}
\small
\centering
\begin{tabular}{lcccccccccccc}
  %\toprule
  \addlinespace[8pt]
  \multirow{1}{*}[11pt]{\textbf{Target}} & 
  \multicolumn{2}{c}{\includegraphics[width=1.cm]{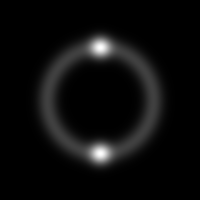}} & 
  \multicolumn{2}{c}{\includegraphics[width=1.cm]{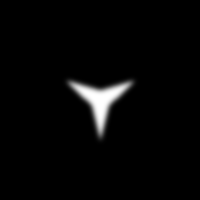}} & 
  \multicolumn{2}{c}{\includegraphics[width=1.cm]{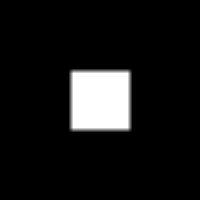}} & 
  \multicolumn{2}{c}{\includegraphics[width=1.cm]{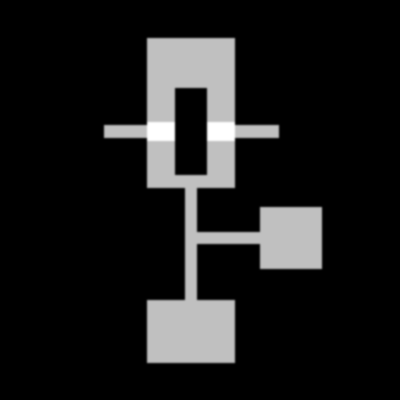}} & 
  \multicolumn{2}{c}{\includegraphics[width=1.cm]{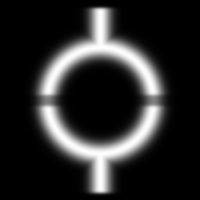}} & 
  \multicolumn{2}{c}{\includegraphics[width=1.cm]{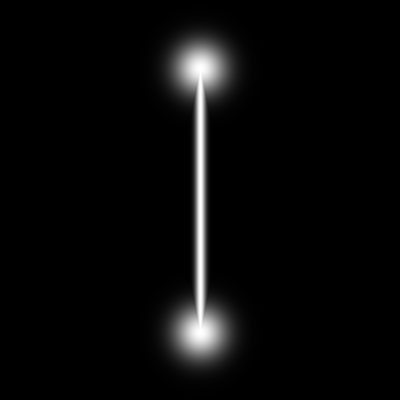}} \\
 \toprule
  Algorithm & $\epsilon$ & $\eta$ & $\epsilon$ & $\eta$ & $\epsilon$ & $\eta$ & $\epsilon$ & $\eta$ & $\epsilon$ & $\eta$ & $\epsilon$ & $\eta$ \\
 \cmidrule(lr){1-1} \cmidrule(lr){2-3} \cmidrule(lr){4-5} \cmidrule(lr){6-7} \cmidrule(lr){8-9} \cmidrule(lr){10-11} \cmidrule(lr){12-13}
  GS~\cite{Pasienski:08} & 21 & 99 & 30 & 99 & 23 & 99 & 34 & 96 & 36 & 97 & 36 & 97 \\
 GS\,+\,OT\,$^a$ & 6.1 & 99.2 & 13.3 & 99.0 & 16.6 & 99.5 & 22.2 & 98.6 & 9.8 & 99.3 & 7.3 & 98.8 \\
 MRAF~\cite{Pasienski:08} & 1.7 & 45 & 2.7 & 29 & 1.5 & 45 & 3.9 & 18 & 1.8 & 30 & 2.9 & 19 \\
 MRAF\,+\,OT\,$^a$ & 0.17 & 72 & 0.96 & 69 & 0.76 & 70 & 0.56 & 53 & 0.35 & 85 & 0.30 & 85 \\
 \bottomrule
 \;$^a$ This work.
\end{tabular}
\end{table*}

\section{Algorithms for phase generation}
\label{sec:otpg}
In this section, we show how optimal transport algorithms can provide approximate solutions to the phase generation problem.  We begin by recapitulating the basic elements of OT theory needed for our work. Detailed treatments can be found in~\cite{Villani2009,Villani2021}. 
 
The basic problem of OT is to find a way of rearranging one probability density $\mu(\mathbf x )$ into another $\nu(\mathbf y )$ that optimizes some cost $c(\mathbf x , \mathbf y)$ for the rearrangement process.  For example, we may think of $\mu(\mathbf x )$ as the height of a pile of sand, $\nu(\mathbf y )$ as the depth of a nearby hole, and $c(\mathbf x , \mathbf y)$ as the cost to move sand from position $\mathbf x$ to fill a hole at position $\mathbf y$.  OT seeks to find a way to move sand into the hole with minimal total cost, encapsulated in a function $\bm \gamma(\mathbf x )$ called the transport map which indicates where to send the sand at location $\mathbf x$, and which minimizes $\int \! c(\mathbf x,\bm \gamma(\mathbf x )) \,\mu(\mathbf x ) \dv{\mathbf x}$. 

The key fact needed from OT theory~\cite{de2014monge,brenier1987decomposition,brenier1991polar} is that in the special case where the probabilities $\mu$ and $\nu$ have domain $\mathbb{R}^n$ and are well-behaved, and where the cost function is $c(\mathbf x , \mathbf y) = \norm{\mathbf x-\mathbf y}_2^2$, an optimal transport map $\bm \gamma$ exists and is the gradient of some scalar function $\phi:\mathbb{R}^n\rightarrow\mathbb{R}$, where $\phi$ satisfies
\begin{equation}
\nu\parens[\big]{\nabla\phi(\mathbf x )}\det\hess\phi(\mathbf x ) = \mu(\mathbf x ).
\label{eq:mae2}
\end{equation}
This is the Monge-Ampere equation (\ref{eq:MongeAmpere}) with $\nu = G^2$ and $\mu = g^2$.  Solving the optimal transport problem with these distributions and the above quadratic cost thus yields an approximate solution to the phase generation problem. 

There is an alternative way of formulating OT problems which is more convenient for computational methods, in which the probability mass $\mu(\mathbf x)$ at a point $\mathbf x$ is allowed to be sent to multiple points of the distribution $\nu$.  In this formulation, the transport map $\bm \gamma(\mathbf x)$ is replaced by a ``transport plan'' $\Gamma(\mathbf x, \mathbf y)$, which is a probability distribution on the product of the domains of $\mu$ and $\nu$.  $\Gamma$ must satisfy $\int \Gamma(\mathbf x,\mathbf y) \dv{\mathbf y} = \mu(\mathbf x)$ and $\int \Gamma(\mathbf x, \mathbf y) \dv{\mathbf x} = \nu(\mathbf y)$, and the value $\Gamma(\mathbf x, \mathbf y)$ is interpreted as how much of the probability mass from point $\mathbf x$ is sent to point $\mathbf y$. 

Efficient computational OT solvers, such as those available via Python~\cite{POT} and Julia~\cite{OptimalTransport.jl} packages, accept as input discretized versions of the distributions $\mu$, $\nu$ (represented as 1D arrays) and cost function $c$ (represented as a 2D array). They return a discretized optimal transport plan $\Gamma$ (represented as a 2D array) which minimizes the total cost $\int \Gamma(\mathbf x, \mathbf y) \, c(\mathbf x, \mathbf y) \dv{\mathbf x} \dv{\mathbf y}$.  For the case of quadratic cost function as above, the optimal transport map $\bm \gamma$ may be recovered from the optimal transport plan $\Gamma$ via the relation
\begin{equation}
\bm \gamma(\mathbf x ) = \frac{1}{\mu(\mathbf x )}\int \mathbf y \; \Gamma(\mathbf x , \mathbf y) \dv{\mathbf y}.
\label{eq:plan-to-map}
\end{equation}
As mentioned above, $\bm \gamma(\mathbf x) = \nabla \phi(\mathbf x)$, where $\phi$ is a solution to \Cref{eq:mae2}.  Thus with $\bm \gamma$ in hand we may compute $\phi$ as 
\begin{equation}
\phi(\mathbf x) = \int_{C_{\mathbf x}} \bm \gamma(\mathbf s) \cdot \dv{\mathbf s},
\label{eq:map-to-phi}
\end{equation}
where the integral follows any path $C_{\mathbf x}$ from a chosen reference point $\mathbf x_0$ to $\mathbf x$.

\subsection{Algorithm description}

In an experimental SLM setup, the data for an instance of the phase generation problem consist of 2D arrays $g^2_{jk}$ and $G^2_{JK}$ representing the input and target output beam intensities sampled on the pixels of the SLM and camera, which have coordinates $(x_j,y_k)$ and $(X_J, Y_K)$, respectively.  Note that the discretized cost function and transport plan will be four dimensional arrays (e.g.\ $\Gamma_{jkLM}$). The discretized transport map $\gamma_{jk,w}$ is a three dimensional array, where the index $w$ takes only two values corresponding to the $x$ and $y$ components of the vector $\bm \gamma$ at point $(x_j,y_k)$.
We compute the OT solution via the following steps:\smallskip
\begin{algorithmic}[1]

\State Flatten $g^2_{jk}$, $G^2_{JK}$ to 1D arrays $\mu_j$, $\nu_K$.
\State Pass $\mu_j$, $\nu_K$, and the discretized, flattened quadratic cost matrix
\begin{equation}
c_{jK} = \left(x_{\left(j\% N\right)}-X_{\left(K\% N\right)}\right)^2 + \left(y_{\left\lfloor{j}/{N}\right\rfloor} - Y_{\left\lfloor{K}/{N}\right\rfloor}\right)^2,
\label{eq:cost-function-matrix}
\end{equation}
(where $m\%N$ denotes remainder of $m$ by $N$) to a computational OT solver, returning a matrix $\Gamma'_{jK}$ representing the discretized, flattened OT plan.
\State Reshape $\Gamma'_{jK}$ to a four dimensional array $\Gamma_{jkLM}$.
\State Compute the discretized OT map $\gamma_{jk,w}$ from first moments of $\Gamma_{jkLM}$ via \Cref{eq:plan-to-map}.
\State Integrate $\gamma_{jk,w}$ via \Cref{eq:map-to-phi}, yielding the discretized OT solution $\phi_{jk}$.

\end{algorithmic}\smallskip

For all simulations in this work, we used the ``sinkhorn'' method of the Julia package OptimalTransport.jl~\cite{OptimalTransport.jl} as the OT solver of step 2, with entropic regularization parameter $\epsilon=0.001$.  In step 5, we use an integration path to each point which starts from a reference point near the center of the computational grid and proceeds parallel first to the x-axis and then to the y-axis.  If the input or output distributions are larger than about $150\times 150$ pixels, we crop or downsample to approximately these dimensions to reduce memory requirements for storing the cost matrix and transport plan, and then after step 5 interpolate $\phi_{ij}$ back to the original grid. We note that such interpolation is only possible because this method produces an unwrapped phase.  Further details for each step are given in the Supplement, and source code is available at~\cite{SLMTools}.  

Truncation errors~\cite{trefethen} generically arise in steps 2, 4, and 5 due to finite grid size, non-zero entropic regularization, finite tolerance of the OT solver, etc.  The integration step 5 also involves a choice of integration path, though we find path-dependent variation is generally modest with a first-order quadrature rule (trapezoid rule).  For the purpose of using OT solutions as initializations to iterative phase retrieval algorithms, we can tolerate a moderate amount of error.  Empirically we find that as long as our OT solution is reasonably close, refining with an IFT algorithm yields a high-fidelity solution as described below.

\subsection{Performance}
\label{sec:ot-performance}

Though OT solutions are intrinsically approximate, we find they are close enough to optimal that using them as initializations to an iterative phase retrieval method such as GS or MRAF results in convergence to an accurate and efficient solution of the phase generation problem.  Crucially, the resulting solutions remain vortex free after GS or MRAF iterations. Figure~\ref{fig:pg-comparison} shows a comparison of phases and resulting output beams generated by various combinations of OT, GS, and MRAF.  

Following~\cite{Pasienski:08,Schroff2023}, we quantify performance of phase generation algorithms using error and efficiency metrics.  The RMS error $\epsilon$ measures the normalized average variation between the target intensity $G^2_{IJ}$ and realized output intensity $\tilde G^2_{IJ}$.  It is defined as

\begin{equation}
\label{eq:RMSError-discrete}
\epsilon(G,\tilde G) \coloneq \sqrt{\frac{1}{|U|}\sum_{(J,K)\in U} \frac{\parens*{\hat G_{JK}^2 - \hat{\tilde G}_{JK}^2}^2}{\hat G_{JK}^4}},
\end{equation}
where the ``measure region'' $U$ is a chosen subset of the output grid, $|U|$ denotes the total number of points of $U$, and $\hat G^2, \hat{\tilde G}^2$ are the target and realized output intensities normalized over the region $U$.  Typically Ref.~\cite{Pasienski:08} defines $U \coloneq \{(J,K) \; | \; G_{JK}^2\geq 0.1 \times \max_{LM} G_{LM}^2\}$, i.e.\ the set of indices $J,K$ for which $G_{JK}^2$ attains at least fraction $1/10$ its maximum value.  In this work we will use this definition except where stated otherwise.  See Supplement for details.   

The efficiency $\eta$ measures what fraction of the input light power is diffracted into the vicinity of the target. It is defined with respect to a given region $V$ in the output plane which is supposed to contain all power of the target beam $G$, and takes the form
\begin{equation}
\eta(\tilde G) \coloneq \frac{\sum_{(I,J)\in V} \tilde G^2_{IJ}}{\sum_{(I,J)} \tilde G^2_{IJ}}.
\end{equation} 
We follow conventions of Ref.~\cite{Pasienski:08} in defining $V$.  $V$ coincides with the MRAF signal region, which typically consists of the locus of points within 10 pixels of a pixel for which the target intensity is at least 10\% of its maximum value.  For some of the targets of \Cref{table:pg-comparison} and \Cref{fig:grid-targets} different definitions of the MRAF signal region are used (see Supplement). 

Table~\ref{table:pg-comparison} shows a comparison of GS and MRAF performance with and without OT initialization, where the statistics for no OT initialization are those of Ref.~\cite{Pasienski:08}.  For both GS and MRAF, using OT initialization provides a factor of 1.4 to 10 improvement in accuracy, and for MRAF the efficiency is simultaneously improved by a factor of 1.6 to 4.5.  Figure~\ref{fig:grid-targets} shows the output beams from various combinations of OT, GS, and MRAF. With our method, tt is possible to use a much larger signal region than in Ref.~\cite{Pasienski:08} while maintaining excellent accuracy and efficiency (see Supplement).  The output beams so generated are shown in the last row of \Cref{fig:grid-targets}. 

\begin{figure}[t]
    \centering
    \includegraphics[width=0.7\columnwidth]{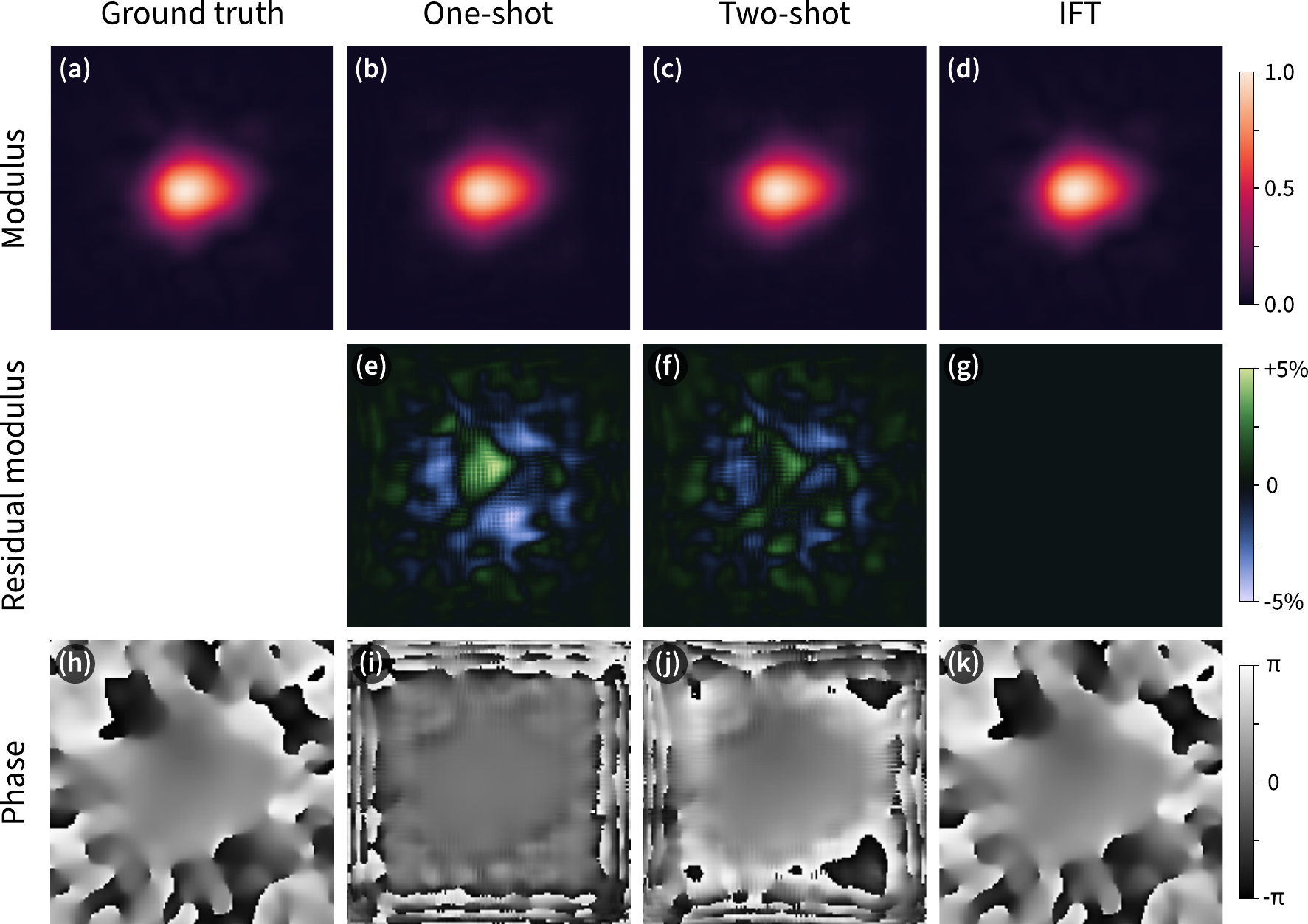}
    \caption{Beam estimates using various phase diversity algorithms.  The top row (a-d) is the beam modulus.  The middle row (e-g) is the residual modulus, i.e.\ the difference between the modulus of the estimate and that of the ground truth.  The bottom row (h-k) is the phase.  (a,h) Ground truth beam. (b,e,i) One-shot beam estimate with diversity coefficient $\alpha=1.5$ ($\delta = 0.02$). (c,f,j) Two-shot beam estimate with diversity coefficients $\alpha_j=1.5$ and $\alpha_k = 0.1$ ($\delta = 0.005$). (d,g,k) IFT estimate with 15 diversity images, $\alpha=0.1,0.2,\dots,1.5$, and 1000 iterations ($\delta = 3.3\times 10^{-17}$). For visual comparison, a global phase has been chosen for each image such that the local phase in the center of the image is $0$.}
    \label{fig:beam-estimates}
\end{figure}

\section{Beam estimation via phase diversity imaging}
\label{sec:phase-diversity}

In this section, we describe algorithms for solving Problem~\ref{prob:beam-estimation} for beam estimation.  The first is an IFT algorithm analogous to GS.  The second uses a single diversity image and the stationary phase approximation of \Cref{eq:spotlight-pdi-phase}.  The third uses optimal transport and \Cref{eq:pd2shot}.  The latter two yield approximate solutions which can be refined by the IFT method. The performance of all three algorithms is discussed at the end of this section.  

\subsection{IFT Phase Diversity Algorithm}
\label{sec:phase-diversity-IFT}
The following algorithm seems to have been introduced first by a patent of Gerchberg~\cite{gerchberg2002system}.  In the Supplement, we show via phase retrieval theory that it is a natural generalization of the GS algorithm to the beam estimation problem.

In the notation of Problem~\ref{prob:beam-estimation}, we are given diversity phase coefficients $\alpha_j$ and corresponding beam moduli $G_j$, $j=1,\dots,n$.  For each index $j$, define a projection operator $P_j$ on the space of complex valued functions $a :\mathbb{R}^2\rightarrow \mathbb{C}$ by
\begin{equation}
\label{eq:pd-proj}
P_j : a \mapsto e^{-2\pi i \, \alpha_j x^2/2}\, \mathcal{F}^{-1}\left[\frac{G_j \, \mathcal{F}\!\left[a\, e^{2\pi i \,\alpha_j x^2/2}\right]}{\left|\mathcal{F}\!\left[a\, e^{2\pi i \,\alpha_j x^2/2}\right]\right|} \right] .
\end{equation}
The set onto which $P_j$ projects is the collection of all complex beam amplitudes $a$ which exactly reproduce the $j$-th diversity image, but not necessarily any other diversity images. 

One iteration of the algorithm is defined to be 
\begin{equation}
    a \gets \frac{1}{n}\sum_{j=1}^{n} P_j(a),
\end{equation}
where $a(\mathbf x) = g(\mathbf x)\, e^{2\pi i \, \psi(\mathbf x)}$ represents the current estimate for the complex input beam and $\gets$ denotes assignment.  In words, at each step we take the current beam estimate, apply the projections for each diversity image, and average the results to get the new beam estimate. 

As with the GS algorithm, iterations are performed either a specified number of times, or until a chosen metric stagnates.  The starting guess for $a$ may either be random or the output of one of the other algorithms below. 

There are several natural variations of this algorithm, such as applying the projections $P_j$ successively rather than averaging them~\cite{almoro2006complete}, or using non-quadratic~\cite{sharma2015phase} diversity phases in Eq. (\ref{eq:pd-proj}).  We have found that the former variant often converges faster, but the error (see \ref{sec:pd-performance}) can cease to decrease monotonically and also depends on the order in which projections are applied.  The latter variant can also lead to stronger convergence~\cite{sharma2015phase}, but we do not consider non-quadratic diversity phases in this work.

\subsection{One-shot beam estimation for collimated beams}
\label{sec:one-shot}
Equation~(\ref{eq:pd-stationary-phase-interp}) asserts that in the case where the input beam has negligible phase, the output beam moduli corresponding to each diversity image are merely rescalings of the input beam modulus.  We may thus estimate the input beam modulus $g$ by interpolation of $G_j$.  Equation~(\ref{eq:pd-one-shot}) achieves essentially the same estimate via a single Fourier transform, with no interpolation required, and is for this reason our preferred method of beam estimation with a single diversity image.  We refer to this as ``one-shot beam estimation''. 

\subsection{Two-shot optimal transport beam estimation}
\label{sec:phase-diversity-OT}
Given two diversity images $j$, $k$, \Cref{eq:pdma} allows us to use optimal transport to solve for the phase $\Phi_j(\mathbf X)$ and thus estimate the input beam by inverse Fourier transformation.  The resulting algorithm is completely analogous to that of \Cref{sec:otpg}, with small modifications to account for the additional phases in \Cref{eq:pdma} vs. \Cref{eq:MongeAmpere} and a final step for the inverse Fourier transform.  The accuracy of this method is controlled by $\alpha_j$ and $\alpha_k$, with better estimates generally coming from one of $\alpha_j$, $\alpha_k$ moderately small and the other as large as practical.  See Supplement for details.  We refer to this method as ``two-shot beam estimation''. 

\begin{figure}[t]
\centering
\includegraphics[width=0.5\columnwidth]{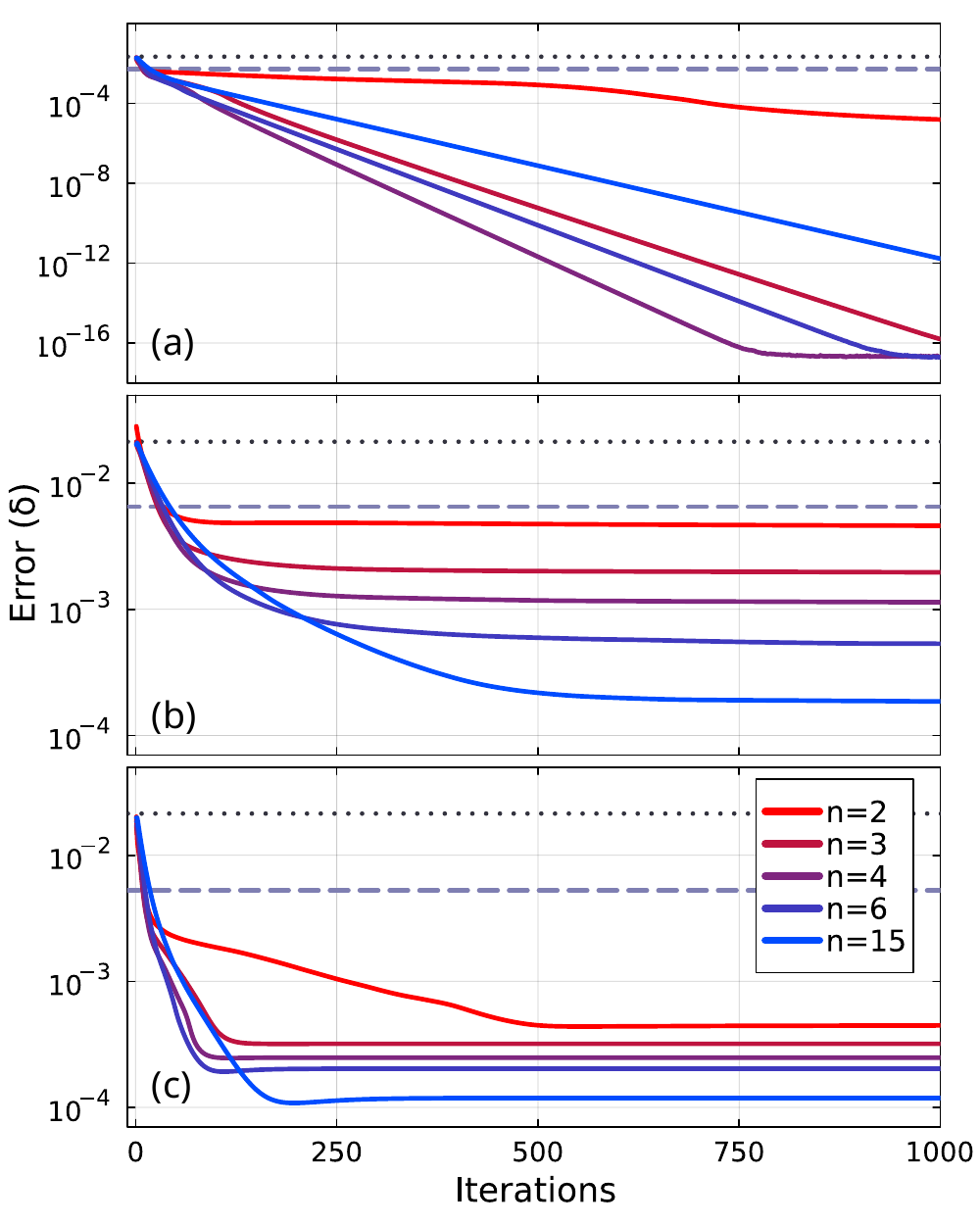}
\caption{Beam estimation error metrics, (a) in absence of noise, (b) with additive noise, and (c) with Poissonian shot noise. Dotted and dashed lines indicate $\delta$ for the one-shot and two-shot algorithms, respectively. Solid lines show $\delta$ vs. the number of iterations of the IFT algorithm with different numbers $n$ of diversity images.  Coefficients $\alpha$ in each case are as follows. $n=2: \alpha \in \{0.1, 1.5\}$. $n=3: \alpha\in \{0.1,0.8,1.5\}$. $n=4: \alpha \in \{0.1,0.6,1.0,1.5\}$. $n=6: \alpha \in \{0.1,0.4,0.7,0.9,1.2,1.5\}$. $n=15: \alpha \in \{0.1,0.2,\dots,1.5\}$.  In (b), we approximate a 16-bit camera with up to two dark counts per pixel by adding to each diversity image pixel $G^2_{j,LM}$ a random value in the range $[0,2^{-15}\times \max_{PQ} G^2_{j,PQ}]$.  In (c), we approximate shot noise for a 16-bit camera by letting each diversity image pixel value be a Poissonian random variable with mean $2^{16}\times G^2_{j,LM}/\max_{PQ}G^2_{j,PQ}$, where $G^2_{j,LM}$ is the corresponding noiseless pixel value.  In all cases, $\delta$ is computed using all 15 noiseless diversity images.
}
\label{fig:pd-convergence}
\end{figure}

\subsection{Metrics and performance}
\label{sec:pd-performance}

In order to quantify beam estimation performance in terms of experimentally accessible quantities, we define an error metric $\delta$ for a beam estimate $\parens[\big]{g(\mathbf x),\psi(\mathbf x)}$ by 
\begin{equation}
\delta \coloneq \sqrt{\frac{1}{n}\sum_{j=1}^n \norm*{ G_j^2 - \abs*{\mathcal{F}\!\left[g \, e^{2\pi i \left(\psi(\mathbf x) + \alpha_j x^2/2\right)}\right]}^2 }_2^2}\,.
\label{eq:pd-error-metric}
\end{equation}
In words, $\delta$ is the $L^2$ distance between the measured diversity image $G_j^2$ and that predicted by the beam estimate, averaged in quadrature over all diversity images. 

We test performance of beam estimation algorithms on a simulated input beam generated by summing Hermite-Gaussian modes with random amplitudes (see Supplement).  Figure~\ref{fig:beam-estimates} shows a comparison of the ground truth input beam and the estimate of modulus and phase produced by each of the above algorithms.  The one-shot (with $\alpha=1.5$) and two-shot (with $\alpha_j = 1.5$, $\alpha_k = 0.1$) estimates have error metrics $\delta = 0.02$ and $\delta = 0.005$, respectively.

In absence of noise, we find that the IFT algorithm of \Cref{sec:phase-diversity-IFT} converges to within machine precision of ground truth (modulo a global phase) when at least 3 diversity images are used.  The rate of convergence depends on the range of diversity phase coefficients $\alpha_j$ and the number of diversity images.  Using more diversity images does not always lead to more rapid convergence (see Discussion).  The rate of convergence is shown in \Cref{fig:pd-convergence}~(a).  

In the presence of image noise, the IFT phase diversity algorithm no longer exactly reproduces the ground truth solution.  Instead, the error metric stagnates at a level which depends on the magnitude of the noise and the number of diversity images used.  In \Cref{fig:pd-convergence}~(b,c) we show the performance of the same three algorithms in the presence of two models of noise.  In computing the error metric in these cases, we use the uncorrupted images $G^2_j$, since this provides a better measure of how close the estimated beam is to the ground truth.

\section{Discussion}
\label{sec:discussion}
The methods we have introduced above for phase generation and beam estimation have much in common and are both largely complementary to other methods in the literature.  Our OT phase generation method can be viewed as an excellent initialization for iterative algorithms.  Many authors have stressed the importance of the initialization step~\cite{pasienski2011transport,Pasienski:08,Schroff2023}, but state-of-the-art methods typically involve a direct search over several parameters and in some cases hand tuning to avoid vortex formation.  Our implementation has only one hyperparameter (a regularization parameter used by the Sinkhorn optimal transport algorithm), which did not require adjustment for the work in this paper. Moreover, we have shown that initializing GS or MRAF with OT solutions can simultaneously improve accuracy and efficiency by significant margins.  

Our beam estimation methods are compatible with any choice of phase generation algorithm.  Additionally, even if one has an independent means of measuring the input beam phase and amplitude (e.g.\ a Shack-Hartmann sensor), phase diversity imaging still offers a convenient way of determining where the beam is incident upon the SLM, which is an important input for any phase generation algorithm.  

A major practical advantage of all of our methods is that they require only modest computational resources.  In particular, no GPU acceleration is needed, which can be an obstacle to using some CFM methods.  Our OT algorithms are nevertheless highly parallelizable (due in particular to the parallelizability of the Sinkhorn algorithm used as an OT solver~\cite{cuturi2013sinkhorn}), which may allow for real time beam shaping applications.  The most significant limitation of our methods is the potential for high memory requirements for OT algorithms, since the size of the cost matrix and transport plan scales as the fourth power of the linear size of an image.  In practice, this is not a major issue for relatively smooth target intensities due to the ability to rescale and interpolate OT solutions.  However, memory requirements could be prohibitive for targets with both large extent and small feature size.  Such memory constraints could be significantly alleviated via a multiscale OT method as outlined in the Supplement, which would also likely result in a significant speedup.

An interesting observation from our comparison of GS and MRAF, with and without OT initialization, is that even when a GS solution is vortex free, its accuracy is still inferior to that achievable with MRAF.  It is well known that vortices are a primary obstacle to achieving accurate output beams via GS~\cite{SENTHILKUMARAN200543}.  Our results show that even in the absence of vortices, the accuracy attainable by GS remains inferior compared to algorithms like MRAF, which are able to boost accuracy (i.e.\ lower $\epsilon$) at the cost of lowering efficiency.

There are several promising directions for extensions of our beam estimation methods.  First, we chose quadratic diversity phases in part to make the two-shot algorithm possible.  However, the IFT algorithm could use arbitrary diversity phases, and other works have investigated phase diversity imaging with vortex~\cite{sharma2015phase,echeverri2016vortex} or random~\cite{schroff2024rapid} phases.  Second, we have some evidence that the deceleration of IFT convergence when many diversity images are used [see \Cref{fig:pd-convergence}~(a)] can be understood in the fractional Fourier domain as an effect of oversampling of low spatial frequencies.  Applying some form of high-pass filtering may alleviate this effect and lead to better convergence.  Finally, it is interesting to investigate the performance of phase diversity under more realistic noise models in an SLM system.

\section{Conclusion}
\label{sec:conclusion}
We have demonstrated new tools for solving the phase generation and beam estimation problems for laser beam shaping with a spatial light modulator. Our methods have many technical advantages over existing alternatives, are user-friendly, and achieve superior performance in simulation. Our contributions are complementary to other techniques commonly used for laser beam shaping with an SLM. Achieving high accuracy output beams in an experimental setup requires consideration of several non-ideal effects we have neglected here, such as SLM pixel crosstalk. Application of the methods of this paper to an experimental setup will be the subject of future work.  

\begin{backmatter}

\bmsection{Acknowledgments}
This work was supported by the Gordon and Betty Moore Foundation Grant GBMF7945, the NSF QLCI Award No.\ OMA-2016244, and partially supported by the U.S.\ Department of Energy, Office of Science, National Quantum Information Science Research Centers, Superconducting Quantum Materials and Systems Center (SQMS) under the contract No.\ DE-AC02-07CH11359.

\bmsection{Disclosures}
The authors declare no conflicts of interest.

\bmsection{Data availability} No data were generated or analyzed in the presented research.  Code for generating all figures and graphs is available in Ref.~\cite{SLMTools}.

\bmsection{Supplemental document}
See Supplement for supporting content. 

\end{backmatter}

\end{document}